\documentclass[twocolumn,showpacs,preprintnumbers,prb,fleqn,superscriptaddress]{revtex4-1}
\usepackage[english]{babel}
\usepackage[utf8]{inputenc}
\usepackage{SIunits}
\usepackage{graphicx}
\usepackage{dcolumn}
\usepackage{amsmath}
\usepackage{natbib}
\usepackage{bm}
\usepackage{textcomp}
\usepackage{color}

\begin{document}

\title{Resonance effects in the Raman scattering of mono- and few layers MoSe$_2$}

\author{P. Soubelet}
\affiliation{Instituto Balseiro and Centro At\'omico
Bariloche, C.N.E.A., R8402AGP Bariloche, R\'io Negro, Argentina}

\author{A. E. Bruchhausen}
\affiliation{Instituto Balseiro and Centro At\'omico
Bariloche, C.N.E.A., R8402AGP Bariloche, R\'io Negro, Argentina}

\author{A. Fainstein}
\affiliation{Instituto Balseiro and Centro At\'omico
Bariloche, C.N.E.A., R8402AGP Bariloche, R\'io Negro, Argentina}

\author{K. Nogajewski}
\affiliation{LNCMI (CNRS, UJF, UPS, INSA),
BP 166, 38042 Grenoble Cedex 9, France}

\author{C. Faugeras}
\email{clement.faugeras@lncmi.cnrs.fr} \affiliation{LNCMI (CNRS, UJF, UPS, INSA),
BP 166, 38042 Grenoble Cedex 9, France}

\date{\today }

\begin{abstract}

Using resonant Raman scattering spectroscopy with 25 different laser lines, we describe the
Raman scattering spectra of mono- and multi-layers 2H-molybdenum diselenide (MoSe$_2$) as well as
the different resonances affecting the most
pronounced features. For high-energy phonons, both A-
and E- symmetry type phonons present resonances with A and B excitons of MoSe$_2$ together with a marked increase of intensity when exciting at higher energy, close to the C exciton energy. We observe symmetry dependent exciton-phonon coupling affecting 
mainly the low-energy rigid layer phonon modes. The shear mode for multilayer displays a
pronounced resonance with the C exciton while the breathing mode has
an intensity that grows with the excitation laser energy,
indicating a resonance with electronic excitations at energies
higher than that of the C exciton.

\end{abstract}

\maketitle

Layered transition-metal dichalcogenides (MX$_2$) are today the subject of intense studies
because of the possibility to isolate a single monolayer, a two dimensional crystal, out of the bulk material~\cite{Mak2010,Wang2012}.
This family of materials includes semiconducting compounds (MX$_2$ with M=Mo or W and X= S, Se, or Te). 
They offer an interesting platform to explore purely two-dimensional excitons,
 phonons, and to study the interplay between the spin and valley
physics~\cite{Xiao2012,Xu2014}. Monolayers of these materials can be
transferred on adapted substrates for optical or transport
investigations. In contrast to gapless graphene, their band gap of
$1-2$~eV allows for the construction of transistors~\cite{Radi2011} and of photodetectors~\cite{Yin2012}.
Even though the room temperature electronic mobilities they exhibit reach only the level of few hundreds
$cm^2.V^{-1}.s^{-1}$, orders of magnitude below what is achieved in conventional III-V or II-VI semiconductors quantum wells, 
their flexibility and transparency are
extremely appealing for future electronic and optoelectronic
applications. For instance, they are used as building blocks of
van der Waals heterostructures~\cite{Georgiou2012,Geim2013,Withers2015}, which cannot be controlled
without a deep knowledge of the phonon modes and of the electron-phonon
interaction in the individual layers. It is also worth to point out that the photoluminescence (PL) of MX$_2$ monolayers arises from tightly bound
excitons~\cite{He2014,Chernikov2014}, with binding energies close
to $400$~meV, and hence comparable to those usually observed in
molecules or in carbon nanotubes~\cite{Scholes2006,Ugeda2014}.

Raman scattering plays an important role in the rapid development 
of the field of research of 2D crystals as it provides, for most layered materials, a fast and non-invasive 
tool to determine the number of layers of a given specimen~\cite{Lee2010}: 
When the thickness of a MX$_2$ flake is reduced, the energy of some 
optical phonons changes in a way that can be traced using standard techniques.
Resonant Raman scattering is powerful
technique to explore electronic properties of materials. The
resonant enhancement of Raman scattering signals when the laser
energy is tuned to particular values corresponding to electronic
excitations of the investigated system, is a well established tool
to perform, for instance, the spectroscopy of weakly radiative
excitations. This technique has been applied to MoS$_2$ thin
layers, probably the most explored layered MX$_2$ so far, and to
WSe$_2$~\cite{Golasa2014,Delcorro2014,Carvalho2015,Scheuschner2015,Lee2015}.

The Raman spectrum of bulk 2H-MoSe$_2$ has been measured in
1980~\cite{Sekine1980}, allowing for the identification of the
main Raman-active phonon modes of this compound. Since that time, only few studies
reported the Raman spectrum of thin layers of this
material~\cite{Tongay2012,Tonndorf2013,Kumar2014}. Although they provided the 
description of the main phonon modes, a complete characterization of the Raman 
scattering spectrum of thin layers of MoSe$_2$ is
still lacking. The phonon band structure of monolayer MoSe$_{2}$
has recently been calculated by ab-initio
techniques~\cite{Horzum2013}, and this compound appears to be
distinct from other member of the MX$_2$ family. All these materials
present 6 optical and 3 acoustical phonon
branches, but, in contrast to the case of MoS$_{2}$, WS$_{2}$ or
WSe$_{2}$~\cite{Molina2011}, according to these calculations, in MoSe$_2$ the A'$_{1}$ mode is
located between the E' and E'' modes and is not the highest
Raman-active mode at the $\Gamma$ point of the phonon band structure.
The phonon mode of E symmetry around $290$~cm$^{-1}$, measured at
$632.8$~nm, has a much weaker intensity than the A$_1$
mode~\cite{Kumar2014} close to $243$~cm$^{-1}$. The E' mode is split over the whole phonon
Brillouin zone, with a splitting of $\sim 4$~cm$^{-1}$ at the
$\Gamma$ point. These two E' phonons are of longitudinal (LO) and
of transverse character (TO), and their splitting comes from the
polar character of MoSe$_2$.

Resonant studies with a large number of laser lines spanning from the near infrared to the ultraviolet are seldom done, and this is particularly difficult and absent in the literature when low energy excitations are involved. It is such a systematic and comprehensive study that is being reported here, for a model MX$_2$ material (MoSe$_2$) from monolayer to multi-layer structures. We present a resonant Raman scattering investigation of mono- and multilayer 2H-MoSe$_2$ using $25$ different laser excitation lines
from $458$~nm ($2.7$~eV) to $825$~nm ($1.5$~eV), which span the energies of the three
different excitons A, B and C, located close to $1.54$, $1.74$ and $2.61$~eV, respectively~\cite{Li2014}, and including both high and low energy vibrations. In the first part, we focus on monolayer MoSe$_{2}$ (1L-MoSe$_2$), discussing in particular the identification of the main Raman scattering features and their excitation profiles. We then address the case of multilayer MoSe$_{2}$ (NL-MoSe$_2$ where N
is the number of layers) and describe the phonon modes which are only
Raman-active in multi-layers. These modes have been observed in
 MoTe$_2$~\cite{Yamamoto2014,Ruppert2014,Froehlicher2015}, MoS$_2$~\cite{Scheuschner2015}, WSe$_2$~\cite{Luo2013,Terrones2014}, 
 and we propose here their proper identification in MoSe$_2$, based on polarization resolved 
 measurement and on resonant Raman scattering experiments. We present the excitation profiles of the main Raman features,
including the low energy shear and breathing modes in the case of 2L-MoSe$_2$, and provide
evidence, for some of these modes, of an increase of the
scattering efficiency at energies beyond that of the C exciton.

Raman scattering experiments were performed at room
temperature in the backscattering geometry with a triple-grating
spectrometer equipped with a nitrogen cooled charge coupled device
(CCD) camera. Laser excitation was provided by a mixed Ar/Kr laser
(from 458 to 676 nm) and by a Ti:Sapph laser (from 680 nm to 825
nm). Laser lines were filtered with a grating used in the Littrow
configuration. A 50x objective was used to focus the excitation
laser down to a spot of $2$~$\mu$m, with optical power of $0.4$~mW.
Thin flakes of MoSe$_2$ were first exfoliated from bulk MoSe$_2$ purchased
from \textit{HQ graphene} and then deposited on a Si/SiO$_{2}$ substrate with a SiO$_2$
thickness of $90$~nm. Flakes composed of 1 to 8 layers were
identified by their different optical contrast contrast (calibrated with AFM measurements 
on other flakes) and then by low energy Raman
scattering and photoluminescence experiments. Raman scattering
spectra were normalized by the integrated intensity of the silicon
peak and by its resonant excitation wavelength
dependence~\cite{Renucci1975}, and by the effect of interferences in the SiO$_2$ layer~\cite{Li2012} (see appendix). 
The luminescence background in the case of 1L and 2L has been removed 
from the presented spectra. In the following, the intensity of
each mode is normalized to its maximum value observed in the
investigated excitation energy range. We are interested in
the independent evolution of the different modes and not by their
relative intensities. The vibration patterns of MX$_2$ has been
described in many recent reports~\cite{Zhang2013,Zhao2013}. Bulk
2H-MX$_2$ belongs to the $D_{6h}$ point group, while their odd and even few
layers counterparts represent the $D_{3h}$ and $D_{3d}$ point groups,
respectively. This implies a change of the symmetry of the phonon
modes depending on the number of layers. In what follows we
will use the previously introduced
notation~\cite{Scheuschner2015}, indicating the symmetry of the
different phonon modes by a double notation corresponding to the odd
and even number of layers.

\begin{figure}
\includegraphics[width=1\linewidth,angle=0,clip]{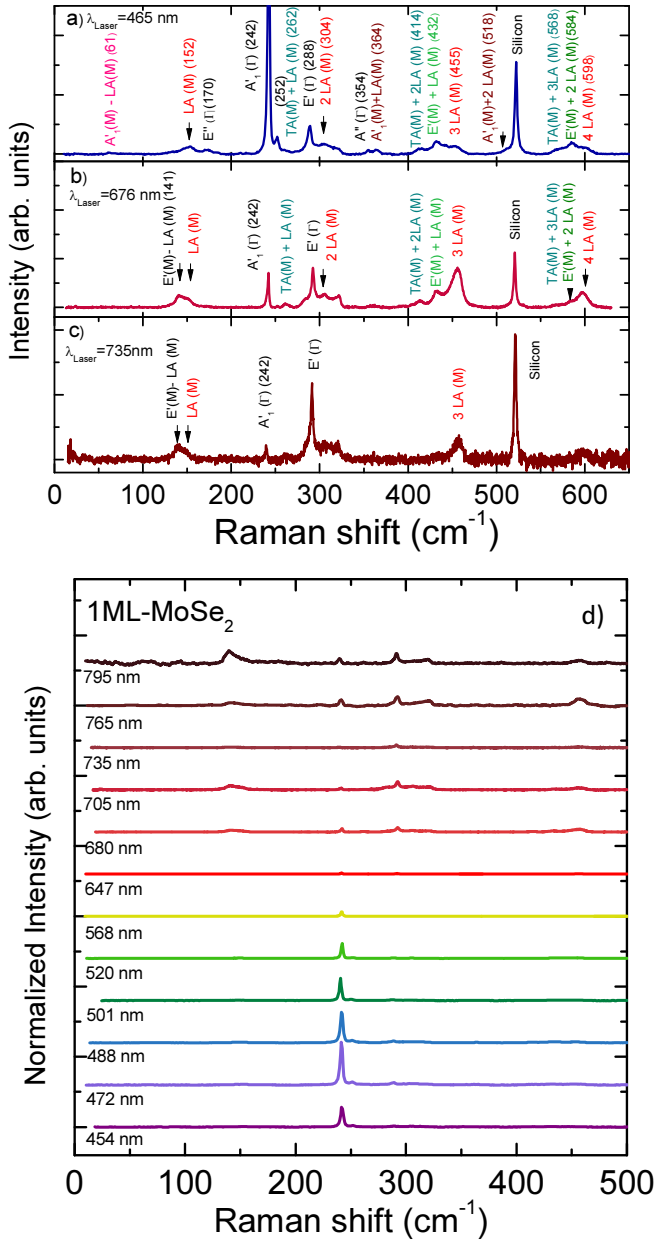}
\caption{\label{Fig1} Raman scattering spectra of monolayer
MoSe$_{2}$ measured with a) $465$~nm, b) $676$~nm and c) $735$~nm laser
excitation. The main Raman scattering features are identified in
the figure and the intensity scale is set in order to see the features for all
excitations wavelengths. d) Normalized Raman scattering spectra for the monolayer.}
\end{figure}

In Fig.~\ref{Fig1}a-c we present three Raman scattering
spectra of monolayer MoSe$_{2}$ measured at $465$~nm (panel a), at $676$~nm (panel b)
and at $735$~nm (panel c). The scale of these spectra is set in order to compare the position of the different features 
and normalized spectra are presented in Fig.~\ref{Fig1}d.
With a $465$~nm excitation, the most prominent
feature is the A'$_{1}$ mode observed at $242$~cm$^{-1}$ in
1L-MoSe$_2$. Its intensity is 10 times bigger
than the intensity of all other Raman scattering features. The E' mode, predicted
to be Raman-active in Ref.~[\onlinecite{Horzum2013}] is also
observed at $288$~cm$^{-1}$. These are the two main features of
the Raman scattering spectrum of
1L-MoSe$_2$~\cite{Tongay2012,Tonndorf2013,Kumar2014,Late2014}.

A more detailed investigation shows that, similar to other MX$_2$~\cite{Berkdemir2013,Chakraborty2013,Golasa2014,Golasa2014b,Lee2015b}
compounds, acoustical phonons also contribute to the Raman
scattering spectrum of 1L-MoSe$_2$
and that depending on the excitation wavelength, such contribution
can be strongly enhanced (see e.g. the spectra measured using $676$~nm and
$735$~nm laser excitations). Because the dispersion of acoustical phonons
 close to the M and K points is rather flat~\cite{Horzum2013}, the
associated density of states is peaked close to this energy and
defect mediated phonon scattering is favored. We observe the longitudinal acoustical
phonon from the M-K points (LA(M)) at $152$~cm$^{-1}$ and its first
overtones (2LA(M), 3LA(M) and 4LA(M)). Next we see a combination
of optical and acoustical phonons from the M point, namely
A'$_1$(M)$\pm$LA(M) at $61$ and $364$~cm$^{-1}$, respectively,
A'$_1$(M)+2LA(M) at $518$~cm$^{-1}$, and E'(M)+LA(M) as well as
E'(M)+2LA(M) at $432$ and $584$~cm$^{-1}$, respectively. One can also notice in our experimental data 
the TA(M)+ XLA(M) with X$=1$ to $3$ at $262$, $414$ and
$568$~cm$^{-1}$. These acoustical phonons 
features are observed on freshly exfoliated MoSe$_2$ flakes, and also, with similar intensity, on flakes 
produced $6$ months before. 

An additional Raman scattering feature appears at $252$~cm$^{-1}$ 
on all samples, mono- or multilayers. Polarization-resolved measurements show
that this feature is of the A-type symmetry, and it follows the same
resonances, discussed in the following, as the main A'$_1$ feature
at $242$~cm$^{-1}$. At high excitation energy, we also observe some
traces of the modes E'' ($\Gamma$) and A''($\Gamma$) at 170 and 354~cm$^{-1}$, 
respectively~\cite{Molina2011,Horzum2013}, which are normally forbidden for mono-layers in back scattering
geometry.

\begin{figure}
\includegraphics[width=0.9\linewidth,angle=0,clip]{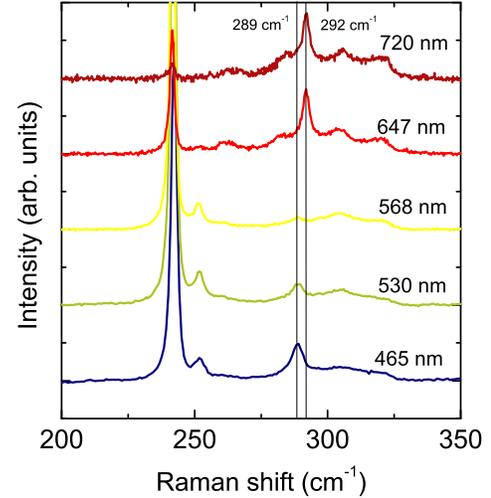}
\caption{\label{Fig2} Raman scattering spectra of monolayer
MoSe$_{2}$ measured with $465$~nm, $530$~nm, $568$~nm, $647$~nm and $720 $~nm
laser excitation showing the selective excitation of the two E'
optical phonons.}
\end{figure}

When decreasing the excitation energy using $676$~nm or $735$~nm excitation (Fig.~\ref{Fig1}b,c), the Raman scattering spectrum of
1L-MoSe$_{2}$ changes drastically: the A'$_{1}$ feature at
$242$~cm$^{-1}$, which is prominent at higher excitation energy now has
an intensity comparable to that of the acoustical phonon features
and of their overtones. The mode TA(M)+LA(M), at
$262$~cm$^{-1}$ and the LA(M) overtones also become particularly
well visible in this excitation energy regime. The E'($\Gamma$) phonon, previously observed at $289$~cm$^{-1}$ now
appears at slightly higher energy $292$~cm$^{-1}$, which we
tentatively interpret, based on the phonon band structure calculations~\cite{Horzum2013}, as the selective excitation, for different excitation 
laser energies, of the lower-energy TO mode or of the upper-energy LO mode~\cite{Horzum2013}. This
crossover is presented in more details in Fig.~\ref{Fig2}. The lower component of the split E' mode,
prominent for excitation wavelengths below $\sim 500$~nm, gradually
disappears as the excitation wavelength increases. Above $\sim
600$~nm, the upper component of the split E' phonon appears at $292$~cm$^{-1}$ with
an intensity increasing with the excitation wavelength. As will be 
confirmed in the following, this behavior suggests a preferential coupling of the 
LO phonon with the A- and B-type excitons, while the TO phonon can only be observed 
when exciting at high-energy, in the C exciton range of energy.
Such an effect, pronounced in 1L-MoSe$_2$, is also observed in
multilayers even though they do not exhibit too strong E'/E$_{g}$ feature for low
excitation energies. We do not have a microscopic explanation for this crossover and for the resonant behavior. 
Electronic band structure calculations, together with a phonon mode analysis might help in understanding this effect. Such analysis are beyond the scope this experimental paper.

\begin{figure}
\includegraphics[width=0.9\linewidth,angle=0,clip]{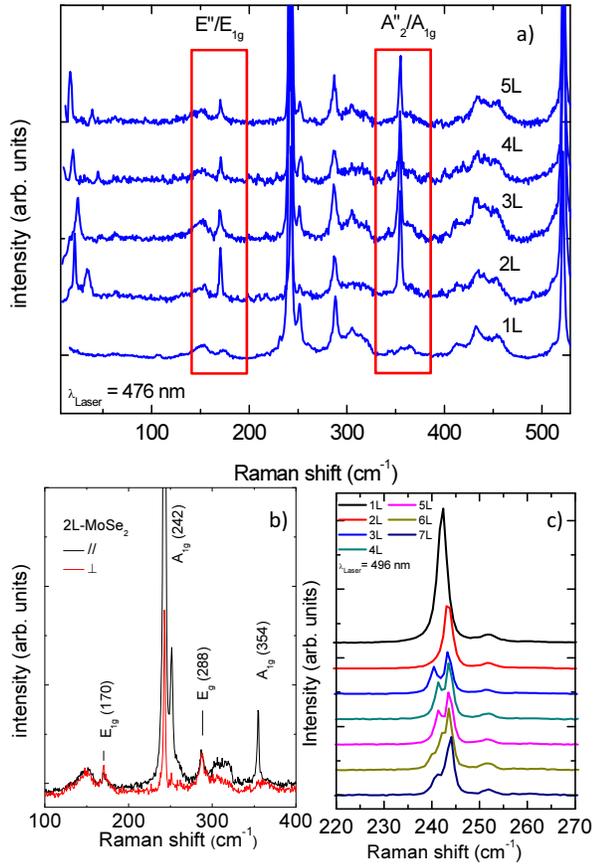}
\caption{\label{Fig3} a) Raman scattering spectra of $1$ to $5$ layer
MoSe$_2$ recorded with the use of $465$~nm excitation showing the E'' and A''$_2$
phonons at $170$ and $354$~cm$^{-1}$, respectively. b) Polarization resolved Raman scattering spectra of 2L-MoSe$_{2}$
measured with $514$~nm and identifying the two E and A type
phonons. c) A'$_1$/A$_{1g}$
feature for 1- to 7- layers MoSe$_{2}$ samples.}
\end{figure}

The Raman scattering spectrum of multilayers of MoSe$_2$ shows
some additional features. As can be seen in Fig.~\ref{Fig3}a for
1L- to 5L-MoSe$_2$ two additional peaks, marked by the red boxes,
appear at $170$ and $354$~cm$^{-1}$ respectively. The energy of
these two modes does not depend on the number of layers. In
Fig.~\ref{Fig3}b, we present two Raman scattering spectra measured
on a 2L-MoSe$_{2}$ in the co- and cross-linear polarization configurations. In
the cross-linear configuration, the A$_{1g}$ mode at
$242$~cm$^{-1}$ is suppressed, together with the mode observed at
$354$~cm$^{-1}$, showing that this mode is of A-type symmetry. The
feature observed at $170$~cm$^{-1}$ is not affected by the
polarization configuration and exhibits hence E-type
symmetry. These two features observed at $170$~cm$^{-1}$ and at
$354$~cm$^{-1}$ are then attributed to the zone center
E''/E$_{1g}$ and A''$_{2}$/A$_{1g}$, respectively, and they
are not Raman-active in monolayers (even though they show up as
weak features when using high excitation energies). Similar Raman
scattering results have been recently obtained on
MoS$_2$~\cite{Scheuschner2015} and the A''$_{2}$/A$_{1g}$ was
observed in WSe$_2$~\cite{Terrones2014}. These latter two modes are not
observed in the MoSe$_2$ monolayer, and, as in the case of MoS$_2$, they only
appear in multilayers for excitation energies close to the
C-exciton energy. The effect have been clearly observed in MoS$_2$~\cite{Scheuschner2015} 
and these authors explained this observation by the fact that, in contrast to the low-energy A and
B excitons whose wave-functions are strongly confined to
individual layers, the C exciton's wave-function is delocalized along the whole
structure~\cite{Qiu2013,Bradley2015}. As a results, the C exciton can
effectively couple to those phonon modes which are characteristic
of multilayers and a resonance of the scattered intensity is
observed for laser energies close to the C-exciton energy. On the other hand, 
features of this type do not appear at lower excitation energies,
corresponding to the A and B excitons~\cite{Scheuschner2015}. The observation of these
two Raman features can help to quickly discriminate between mono-
and multi-layer MoSe$_2$ flakes. 

A more detailed knowledge of the number of layers constituting the sample under investigation 
from Raman scattering experiments can be grasped
from the observation of the A'$_1$/A$_{1g}$ feature close to
$242$~cm$^{-1}$ (see Fig.~\ref{Fig3}c). Indeed, as
already reported~\cite{Tonndorf2013}, this peak has an energy that depends on the
number of layers, appears in the Raman scattering spectrum as a single component feature for 1L-MoSe$_2$ and
2L-MoSe$_2$, splits into two components for 3L-MoSe$_2$ and
4L-MoSe$_2$ and into three components for 5L-MoSe$_2$ and
6L-MoSe$_2$. The splitting, which comes from interlayer interactions in multilayer 
materials ($N>2$)~\cite{Verble1970,Sekine1980,Froehlicher2015}, can be used to
determine the number of layers. Close to $252$~cm$^{-1}$, another Raman scattering 
feature is observed with an energy that changes with the number of layer. Even though 
the proper origin of this feature is still unknown, its energy difference with the 
A'$_1$/A$_{1g}$ feature can provide an additional check for the number of layers of MoSe$_2$ 
(see appendix). In NL-MoSe$_2$, the contribution
of acoustical phonons is also observed in the Raman scattering
spectrum in the form of a rich series of overtones, as already discussed in
the case of the monolayer.

To complete the description of the differences between mono-
and multilayers, we note that another striking difference is seen
at low vibrational frequencies, below $50$~cm$^{-1}$. In multilayers, as can be observed in Fig.~\ref{Fig3}a, some
additional features appear in that range with an energy that strongly depends on
the number of layers. These modes are the shear and breathing modes
of the multilayer structure, i.e. rigid vibrations of the whole layers with
respect to each other. The evolution of these modes
with increasing number of layers has recently been discussed in
the case of MoSe$_2$~\cite{Yan2015}. One can notice from the spectra
presented in Fig.~\ref{Fig3}a that, when measured with a $476$~nm
excitation, the intensity of these modes is comparable with that
of the E''/E$_{1g}$ and A''$_{2}$/A$_{1g}$ modes discussed
previously.

\begin{figure}
\includegraphics[width=0.9\linewidth,angle=0,clip]{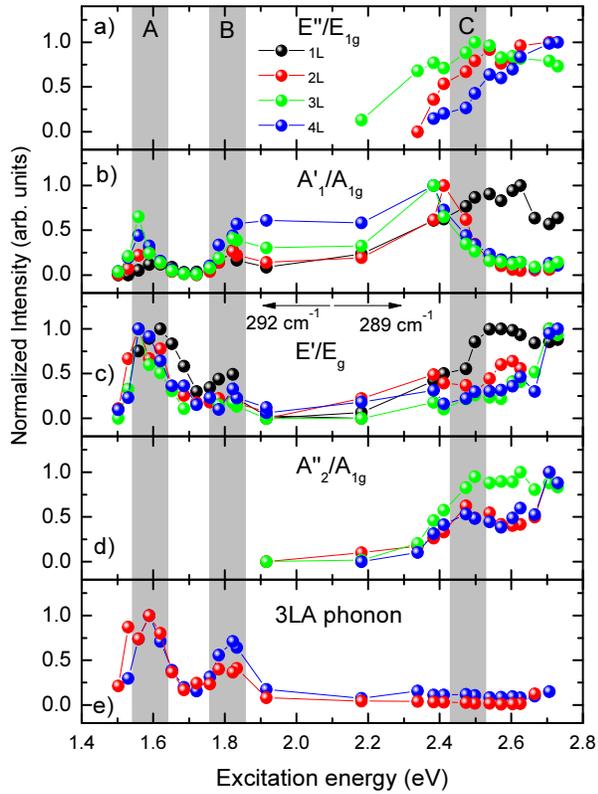}
\caption{\label{Fig4} Normalized intensity of a) the E''/E$_{1g}$
mode at 170 cm$^{-1}$, b) the A'$_{1}$/A$_{1g}$ mode at
$242$~cm$^{-1}$, c) the E'/E$_{g}$ modes around $290$~cm$^{-1}$, d) the A''/A$_{1g}$ modes at $354$~cm$^{-1}$, 
and the 3LA phonon replica at $455$~cm$^{-1}$, as a function of
the excitation laser energy for 1L-MoSe$_{2}$ (black dots),
2L-MoSe$_{2}$ (red dots), 3L-MoSe$_{2}$ (green dots) and
4L-MoSe$_{2}$ (blue dots). The shaded regions represents the range of energy for the A, B and C excitons.}
\end{figure}

When changing the excitation laser energy, all the Raman scattering features considered so far show particular resonances.
These resonances for one- to four- layer flakes, are presented in
Fig.~\ref{Fig4} for the E''/E$_{1g}$ phonon at 170~cm$^{-1}$ (panel a),
the A'$_{1}$/A$_{1g}$ phonon close to 242 cm$^{-1}$ (panel b), the E'/E$_{g}$ phonon close
to $290$~cm$^{-1}$ (panel c), for the A''$_2$/A$_{1g}$ phonon at 354~cm$^{-1}$ (panel d), 
and for the 3LA(M) replica (panel d). The E''/E$_{1g}$ and
A''$_2$/A$_{1g}$ Raman modes at 170~cm$^{-1}$ and at
354~cm$^{-1}$, respectively, typical for multilayers, are only observed for excitation
energies above $2.2$~eV. They both exhibit a resonance close to
$2.5$~eV. For mono- and multilayers, we observe a resonance 
of the LO E'/E$_{g}$ phonon ($\sim 292$~cm$^{-1}$) with the A and B excitons at $1.6$ and $1.8$~eV, respectively, while the TO E'/E$_{g}$ 
phonon ($\sim 289$~cm$^{-1}$) has an intensity that increases when exciting at higher energy,close to the C excitons at $2.4$~eV. As a results, in our experiments, these two LO and TO phonons cannot be observed simultaneously.
On top of these resonances, the intensity of the
TO E'/E$_{g}$ mode appears to continuously increase when increasing
the excitation laser energy. In the case of the monolayer, for
which these two features are the most pronounced, the resonance
appears at a slightly higher excitation energy, close to $2.6$~eV.

The main A'$_1$/A$_{1g}$ feature close to 242 cm$^{-1}$
(Fig.~\ref{Fig4}b) is observed for all samples and shows for the 1L-MoSe$_2$ a
resonance at high energy close to $2.6$~eV,
which corresponds to the C-exciton energy~\cite{Li2014}. For higher
number of layers, this high energy resonance appears at an energy
close to $2.4$~eV. A second resonance is observed at $1.8$~eV and
corresponds to the B exciton. Within the resolution of our
experiment, the energy of the resonance does not seem to depend
on the number of layers. Lowering further the excitation laser energy, 
another resonance is observed. As for the LO E'/E$_{g}$
feature, this resonance occurs close to $1.6$~eV and its energy 
strongly changes with the number of
layers. The energies of the A- and of the C- exciton energies appear to depend 
on the number of layers while the energy of the B excitons does not. 
Such evolution for the A, B and C excitons as a function of
the number of layers has recently been observed in WSe$_{2}$ and in MoSe$_{2}$~\cite{Arora2015,AroraMoSe}. 
These results, concerning the various resonances observed for all the phonon modes of
MoSe$_2$, appear as different as compared to the ones reported for MoS$_2$, where
the A-type phonon mode shows a resonance with the high-energy
exciton only while the E-type modes show a resonance with exclusively the
low-energy A and B excitons~\cite{Carvalho2015}. In MoSe$_2$, the A'$_1$/A$_{1g}$ 
mode shows clear resonances with all three types of excitons, while the E'/E$_{g}$ 
LO phonon displays a resonance with the A and B excitons and the E'/E$_{g}$ TO 
phonon only appears at high excitation energy. 

Some modes, namely the E'/E$_{g}$ and the A''$_2$/A$_{1g}$, have an intensity that, on top of the resonances observed at the exciton energies, increases strongly when increasing the excitation laser energy. This second resonance could be related to the valence band, and especially the band located $\sim 2$~eV below the highest valence band at the K point~\cite{Horzum2013,Ugeda2014,Wang2015a}. In this range of energy, the partial density of states shows a maximum that could explain the resonance behavior observed in Raman scattering. This hypothesis needs to be further clarified on a theoretical point of view and/or with Raman scattering experiments performed in the UV range of energy.
Finally, in contrast to the case of optical phonon modes described above, the LA phonon replicas show a pronounced resonance only with the low energy A and B modes, and are only weakly visible when exciting at higher energy. This behavior is shown in Fig.~\ref{Fig4}e for the 2L and 4L-MoSe$_2$. Similar to the case of MoS$_2$~\cite{Golasa2014}, when using an excitation energy close to that of A or B excitons, these acoustical phonon replica can become the dominant contribution to Raman scattering spectrum of MoSe$_2$ (see spectrum at $676$~nm in Fig.~\ref{Fig1} and appendix).  

\begin{figure}
\includegraphics[width=0.9\linewidth,angle=0,clip]{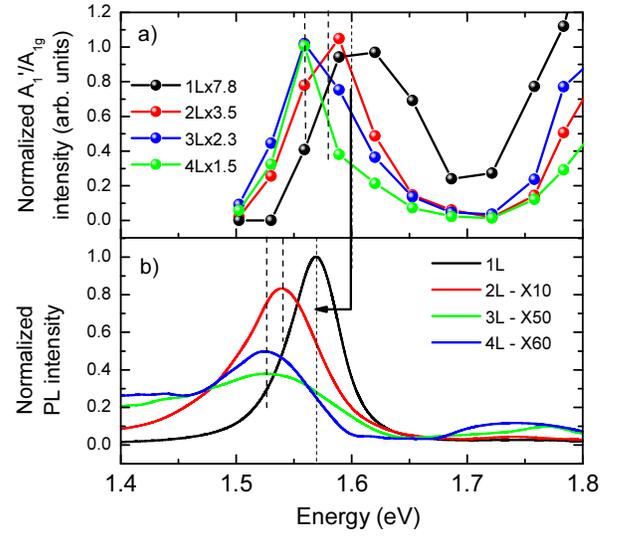}
\caption{\label{Fig5} a) Normalized intensity of the A$_{1}$ mode
at $242$~cm$^{-1}$ as a function of the excitation laser energy.
b) Photoluminescence intensity, normalized to the maximum
intensity of the monolayer. Vertical dashed lines indicate the observed maxima of Raman intensities (panel a) and of luminescence (panel b). Note the shift between the two maxima (highlighted with the vertical arrow, and equal to the energy of the involved phonon), indicative of an outgoing resonance process.}
\end{figure}

The low-energy resonance of the A'$_1$/A$_{1g}$ mode is presented
in more detail in Fig.~\ref{Fig5}a for excitation energies
between 1.4 and 1.8 eV. The photoluminescence (PL) of MoSe$_2$
mono-layers arises from the low-energy A
exciton~\cite{MacNeill2015}. In Fig.~\ref{Fig5}b, the room-temperature PL spectra recorded 
on the same samples are plotted for comparison with the data displayed in Fig.~\ref{Fig5}a.

When increasing the number of layers, we observe at room temperature a marked decrease of the luminescence intensity together with a small shift of the luminescence energy, from $1.57$~eV for 1L-MoSe$_2$, to
$1.54$~eV for 2L-MoSe$_2$, and to $1.52$~eV for 3L- and 4L-MoSe$_2$.
The origin of these two effects is still debated. They could originate (i) from a direct to indirect band gap transition when going from mono to multilayers, or (ii) reflect the difference in the relative energies of the A exciton and of the single particle indirect bandgap due to the change of dielectric screening and the decrease of the exciton binding energy for multilayers.
The observed resonances for the A'$_1$/A$_{1g}$ Raman scattering
feature are centered at $1.610\pm0.01$, $1.575\pm0.003$,
$1.560\pm0.002$ and $1.560\pm0.002$~eV for 1L- to 4L-MoSe$_2$,
respectively. We interpret the energy difference between the 
emission of A excitons and the resonance in the Raman scattering as a 
signature of an efficient outgoing resonance. In fact, the maximum intensity 
of the Raman scattering is observed in this material when the incoming photon 
energy is equal to the sum of the exciton and of the scattered phonon energies.

\begin{figure}
\includegraphics[width=0.9\linewidth,angle=0,clip]{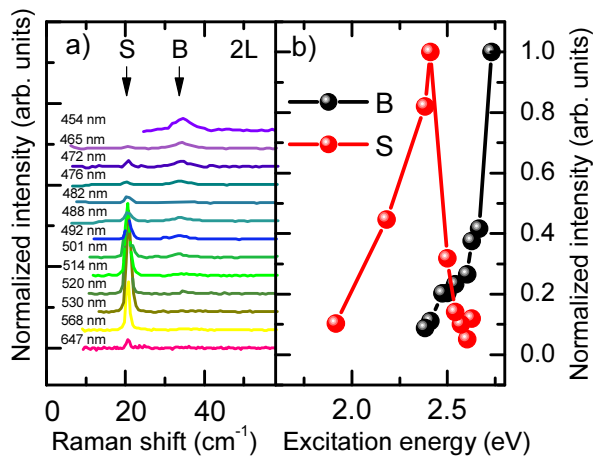}
\caption{\label{Fig6} a) Raman scattering spectra of 2L-MoSe$_{2}$
for excitations energy ranging from $458$~nm to $647$~nm showing
the shear (S) and breathing (B) modes. b) Evolution of the
normalized intensity of the shear mode (red dots) and of the
breathing mode (black dots) as a function of the excitation laser
energy.}
\end{figure}

As already mentioned, layered materials also show low-energy shear phonon modes, of the 
E-type symmetry, and breathing phonon modes, of the A-type symmetry.
Detailed investigation of these modes in few-layer samples has
recently become more accessible due to the commercial development of volume
Bragg grating filters allowing to easily explore the low-energy 
range~\cite{Tan2012} of the Raman scattering spectrum, down to 5-10 cm$^{-1}$. 
The shear and breathing modes are today the subject of intense research in the field of MX$_2$. 
Their Raman shift as a function of the number of layers and their polarization selection
rules have been described in detail for MoS$_2$, WSe$_2$,
MoSe$_2$, WS$_2$ and in MoTe$_2$~\cite{Zhang2013,Zhao2013,Yan2015,Froehlicher2015}. Because of technical difficulties, such studies are usually performed at a single excitation wavelength, and a complete picture of the excitation 
profile of such low energy modes in MX$_2$ is still lacking. Resonant
Raman spectroscopy of such low-energy modes has been performed in 
semi-metallic twisted graphene, allowing to trace the evolution of its high-energy
bands and to quantify the interlayer interaction as a function of
the twist angle~\cite{Wu2014}.

The use of a triple spectrometer and the access to large number of laser lines has allowed us to perform such a detailed resonant study for mono and multilayer MoSe$_2$. In Fig.~\ref{Fig6}a, we present Raman scattering spectra of the shear (S) and breathing (B) modes
of 2L-MoSe$_2$ measured for excitation wavelengths from $454$~nm to $647$~nm.
They are observed at $21$~cm$^{-1}$ and $34$~cm$^{-1}$, respectively~\cite{Yan2015} and their energies do 
not change with the excitation wavelength. Fig.~\ref{Fig6}b illustrates the evolution of the normalized intensity of these two
modes as a function of the excitation laser energy. As can be seen, they show rather distinct resonance behaviors. 
The E-type symmetry shear mode exhibits a pronounced resonance close to
$\sim 2.4$~eV, similar to the resonance observed for both high-energy A-type 
modes, which corresponds to the C exciton energy
in 2L-MoSe$_{2}$. In contrast, the breathing mode, of the A-type symmetry, 
does not show this resonance but its intensity
increases with increasing excitation laser energy. At excitation energies below $2.4$~eV, 
the intensity of the breathing mode is nearly not observable in 2L-MoSe$_2$, grows rapidly 
at higher energies, becoming stronger than that of the shear mode and eventually 
representing the most pronounced low energy feature above $2.65$~eV. For excitation energies lower 
than $1.95$~eV (647 nm), we could not observe these low-energy
modes mainly because of the strong low-energy scattering and/or PL when 
approaching the energies of the A and B excitons. Similar to the TO E'/E$_g$ 
at $289$~cm$^{-1}$ and to the A''$_{2}$/A$_{1g}$ mode at $354$~cm$^{-1}$, 
the breathing mode's intensity seems to show a resonance with electronic excitations above 
the energy of the C exciton. For different MX$_2$ materials, depending on the excitation 
laser and on the exciton energy structure, very different intensity ratios between 
the shear and breath modes have been reported recently~\cite{Zhang2013,Zhao2013,Yan2015,Lee2015,Froehlicher2015}.
As follows from the study reported here, these differences arise from the 
peculiar resonance effect affecting differently the two modes.

To conclude, we have presented a comprehensive Raman scattering
study of MoSe$_2$ mono- and multi-layers. We have shown in this compound 
the existence of new Raman modes which are only active in multilayers, and described their
resonances when tuning the excitation laser energy
across the three different excitons of this material. Resonant Raman
spectroscopy appears as a well adapted tool to study a variety of
excitons in thin layers of MX$_2$ and their coupling to optical
phonons. Our experiments confirm the lowering of the A exciton's
energy when increasing the number of layers and this result is
corroborated by the PL measurements. We also show symmetry-dependent exciton-phonon coupling, in particular, we show
experimentally that the low-energy shear mode of a bilayer displays a
a pronounced resonance with the C exciton while the breathing mode
does not. On the other hand, this latter mode, together with the
E'/E$_g$ and A''$_2$/A$_{1g}$ mode have an intensity that grows
with the excitation laser energy, which is a signature of a still
unexplored resonance at higher energy, in the deep UV range.

\begin{acknowledgements}
We acknowledge fruitful discussions with S. Berciaud, M. Molas and with M.
Potemski. Part of this work has been supported by the TWINFUSYON project, the graphene
flagship project and by the European Research Council
(ERC-2012-AdG-320590-MOMB).
\end{acknowledgements}

\section{Appendix}

\subsection{Photograph of the different MoSe$_2$ specimens}

\begin{figure}
\includegraphics[width=0.7\linewidth,angle=0,clip]{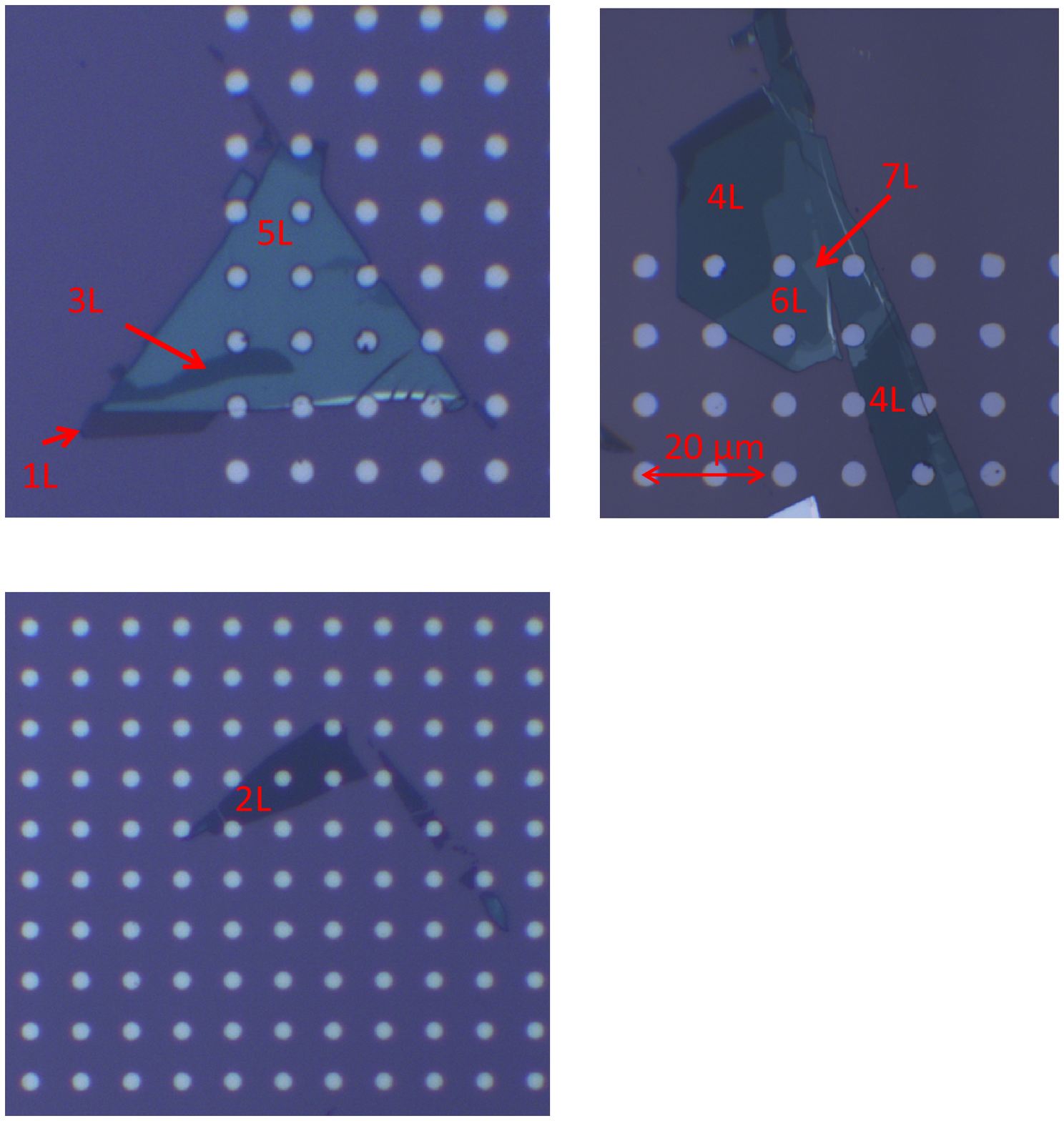}
\caption{\label{Samples} Three optical photographs of the same wafer with different locations corresponding to 1L, 2L, 3 L, 4 L, 5L and 6 L MoSe$_2$ specimens.}
\end{figure}

Large flakes of MoSe$_2$ mono- and multi-layers have been produced by mechanical exfoliation of bulk MoSe$_2$. Flakes were then deposited on a Si/SiO$_2$ substrate with $90$~nm of SiO$_2$. The substrate had been patterned with circular holes of $6$~$\mu m$ diameter, but all the results presented in the main text have been obtained on supported regions. The following photograph in Fig.~\ref{Samples} present the different specimens that were used for the reported measurements.

\subsection{Raman scattering features at $242$ and $252$~cm$^{-1}$}

\begin{figure}
\includegraphics[width=0.7\linewidth,angle=0,clip]{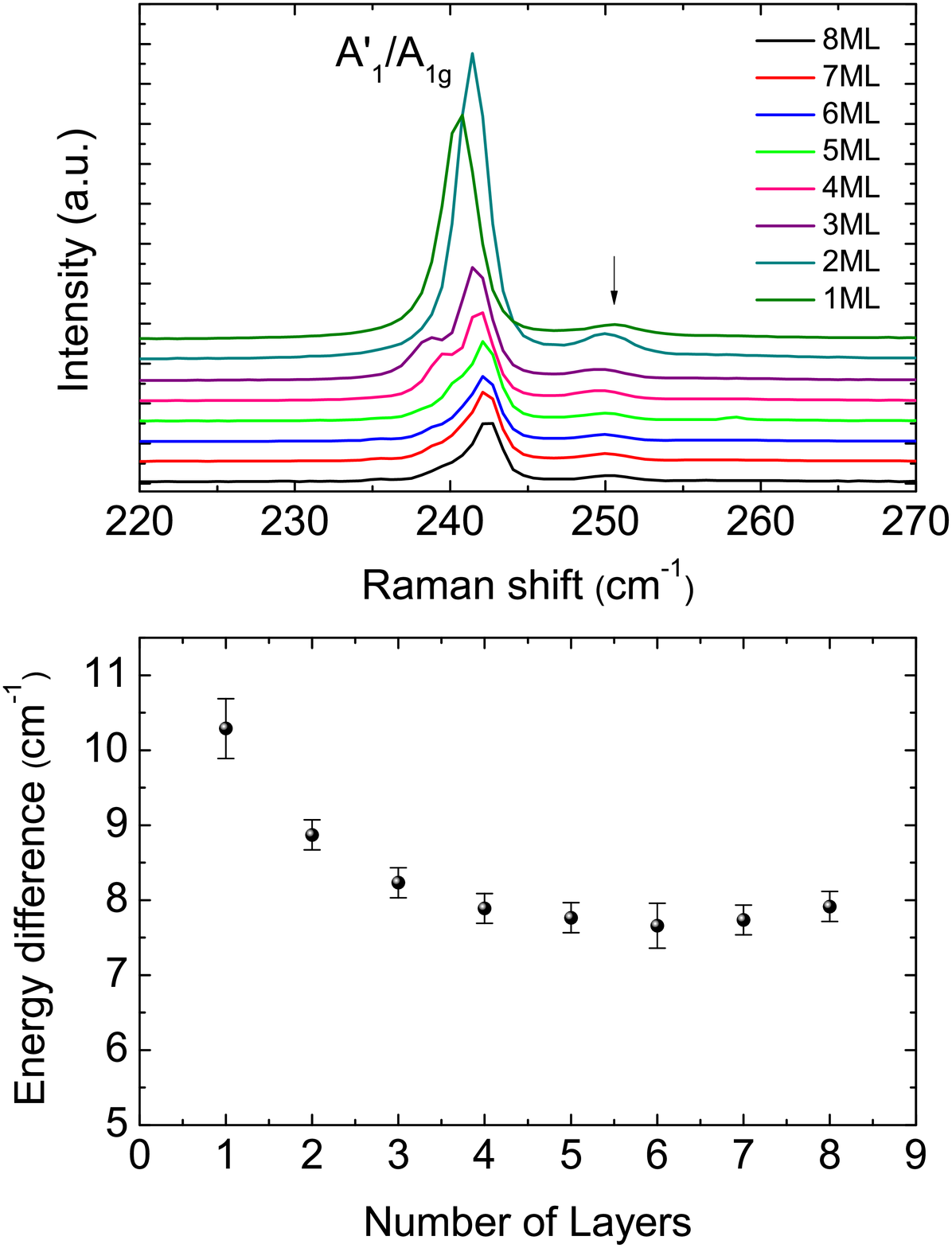}
\caption{\label{Dif} Upper panel: Raman scattering spectra for 1L to 8L-MoSe$_2$ samples. Lower panel: Energy difference between the $252$~cm$^{-1}$ feature and the A'$_1$/A$_{1g}$ feature. The energy separation between these modes is maximum in the case of the monolayer and then decreases for thicker samples. The arrow in the top panel indicates the Raman scattering feature observed close to $250$~cm$^{-1}$.}
\end{figure}

As stated in the main text, MoSe$_2$ mono-layers can be identified by (i) a single component A'$_1$ feature at $240$~cm$^{-1}$ while this feature is observed at $242.5$~cm$^{-1}$ for 2L and is a multi-component feature for $N>2$, (ii) the non-observation of the two E''/E$_{1g}$ and A''$_{2}$/A$_{1g}$ modes at $170$ and $354$~cm$^{-1}$, respectively, and (iii) by the observation of an energy separation bigger than $9$~cm$^{-1}$ between the A'$_1$ and the feature close to $252$~cm$^{-1}$ (see Fig.~\ref{Dif}. These ar the three Raman scattering signatures of 1L-MoSe$_2$ that emerge from this study.

\subsection{Spectra normalization procedure}

The normalization of Raman scattering intensities is crucial for the study of resonance effects. All intensities presented in the main text are normalized according to the following procedure:
We first make a background correction. This background is most of the case flat but, when the excitation wavelength is set close to the low energy excitons of mono or few layer MoSe$_2$, luminescence signals affect the spectra and are removed but using standard computer tools.
Spectra are then normalized by the integrated silicon signal at $522$~cm$^{-1}$.
The silicon first order Raman scattering peak shows a resonance when using high excitation energies. This effect was studied in detail in Ref.~[\onlinecite{Renucci1975}] and has to be taken into account by normalizing the measured signal by the resonance curve of silicon (which also includes the wavelength dependance of the Raman scattering process). Finally, the flake is on top of a SiO$_2$/Si structure which produces optical interferences which have a strong impact on the measured Raman intensities. As it is shown in the literature~\cite{Yoon2009,Li2012,Carvalho2015}, the Raman intensities of the Si and the MoSe$_2$ layers are proportional to an \textit{enhancement factor} due to the optical interference existent in a multi-layered structure given by:
\begin{equation}
 I_{MoSe_2} = \int_0^{d} |E_{ex}(x)E_{sc}(x)|^2 dx,
\end{equation}
where $E_{ex}(x)$ and $E_{sc}(x)$ are the electrical field's amplitude of the total incident and scattered light respectively, and $d$ is the thickness of the MoSe$_2$. The enhancement factor for the Si substrate is~\cite{Li2012,Carvalho2015}
\begin{equation}
\label{ec:dos}
 I_{Si} = \int_0^{\infty} |F_{ex}(x)F_{sc}(x)|^2 dx,
\end{equation}
where $F_{ex}(x)$ and $F_{sc}(x)$ are the electrical field amplitude of the total incident and scattered light respectively. It is important to take into account the wavelength dependence of the complex refractive index of each materials to describe the propagation of the incident and of the scattered light. In the experiment, the incident wavelength is given by the exciting laser and the Raman process defines the scattered wavelength. To correct the data, we multiply the experimental spectra by the ratio $I_{Si}/I_{MoSe_2}$~\cite{Li2012,Carvalho2015} after performing the silicon signal normalization described above.

\begin{figure}
\includegraphics[width=1\linewidth,angle=0,clip]{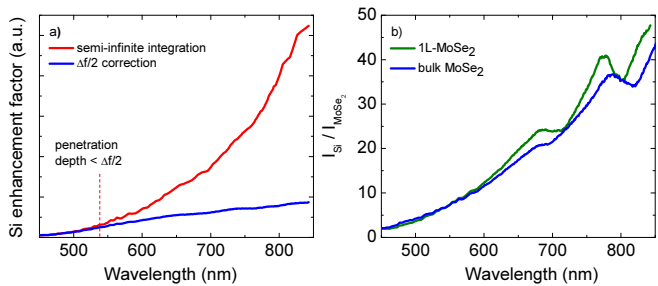}
\caption{a) Interference enhancement factor for the Si Raman signal ($I_{Si}$) as a function of wavelength for a semi-infinite integration and the $\Delta f /2$ correction. When the penetration depth is smaller than $\Delta f /2$, (400 - 550\,nm) the $I_{Si}$ factor is approximately the same for both calculations. b) Ratio $I_{Si}/I_{MoSe_2}$ for 1L-MoSe$_2$ and for bulk MoSe$_2$.\label{Pedro}}
\end{figure}

Expression \eqref{ec:dos} is calculated considering an ``unidimensional'' problem, where the Si substrate is treated as a half-infinite media. In the IR region, where the Si absorption is small, when using a microscope objective with a large numerical aperture, we should take the divergence effect of the focused gaussian beam into account and calculate the integral in a probed volume defined by the depth of focus. The depth of focus $\Delta f$ is defined as the distance between the points at each side of the beam-waist where the irradiance of the beam falls to one-half of its value~\cite{Delhaye1996}. According to Ref.~[\onlinecite{Delhaye1996}] it is given by

\begin{equation}
\label{eqf}
  \Delta f = \frac{\pi(2w_0)^2}{2\lambda},
\end{equation}

where $w_0$ is the beam radius at the waist (spot size in the focus) and $\lambda$ is the wavelength within the material. Taking that into account, the upper integral limit($\infty$) must be modified to $\Delta f/2$.

For 1L-MoSe$_2$ sample, Fig.~\ref{Pedro}a shows the comparison between the semi-infinite integration and the $\Delta f /2$ correction for the Si Raman intensity both as a function of wavelength. Here we assumed a Raman shift of 520\,cm$^{-1}$. The complex refractive index for Si and SiO$_2$ as a function of the wavelength were taken from Ref.~[\onlinecite{Palik1998}] and Ref.~[\onlinecite{Malitson1965}] respectively and the refractive index of 1L-MoSe$_2$ was taken from Ref.~[\onlinecite{Li2014}]. Fig.~\ref{Pedro}b shows the ratio of the Si and the MoSe$_2$ enhancement factor as a function of the wavelength for a 1L-MoSe$_2$ and bulk MoSe$_2$ assuming a Raman shift of 290\,cm$^{-1}$. The bulk MoSe$_2$ refractive index was taken from Ref.~[\onlinecite{Li2012}]. As can be seen, the enhancement factor for 1L-MoSe$_2$ and bulk MoSe$_2$ are similar so that we used the refractive index of the 1L-MoSe$_2$ for the calculations in the multi-layer samples.

\end{document}